# Dodging the electricity price hike: Can demand-side flexibility compensate for spot price increases for households in Germany?

JUDITH STUTE[1], SABINE PELKA[2], MATTHIAS KÜHNBACH[3], MARIAN KLOBASA[2]

[1]Fraunhofer Research Institution for Energy Infrastructure and Geothermal Systems IEG, Breslauer Str. 48, 76139 Karlsruhe, Germany, judith.stute@ieg.fraunhofer.de, corresponding author

[2]Fraunhofer Institute for Systems and Innovation Research ISI, Breslauer Str. 48, 76139 Karlsruhe, Germany

[3]Fraunhofer Institute for Solar Energy Systems ISE, Heidenhofstr. 2, 79910 Freiburg, Germany

KEYWORDS

Dynamic tariffs; electric vehicle; heat pump; energy management system; smart meter; demand response

ABSTRACT

In 2022, energy prices skyrocketed across Europe. Among other things, average electricity prices on the day-ahead spot market in Germany were 2.43 times higher than the year before, which may be a preview of future prices. At the same time, electricity infrastructure is expected to be overutilized in some regions in the future due to the uptake of electric vehicles, heat pumps, PV systems, and other appliances in the residential sector. One option discussed for easing the burden on the infrastructure and reducing electricity costs for households is to have dynamic electricity prices for residential customers. Time-varying prices over the course of the day and year allow these customers to profit from operating their assets more flexibly. The question arises as to what impact dynamic electricity prices as recently seen on the day-ahead spot market in Germany would have on the use of flexibility in households and whether it can compensate for cost increases for households. We analyze flexibility utilization to (a) take advantage of dynamic electricity prices and compare the cost savings with additional costs for installing a home energy management system (HEMS) and smart meters and (b) only increase self-consumption of PV systems using a static tariff and compare the outcome with results of (a). We show that with higher electricity prices and price spreads, a higher share of households can achieve cost savings by utilizing flexibility from their electric vehicles, heat pumps, and battery storage systems and compensate initial investments for HEMS and metering point operation costs. Under current conditions using flexibility to increase self-consumption remains an attractive option for households with a heat pump if a PV system or a PV-battery storage system is available (considering also costs for HEMS and smart meters). For households with electric vehicles, dynamic tariffs are financially more attractive than solely focusing on self-consumption. For the pricing scenarios studied, we determine how high the annual additional costs incurred by investments in a HEMS and metering point operation costs can be to still achieve cost savings for 75% of the households analyzed and thus enable the utilization of flexibility. Results show, that in the self-consumption case, those costs range from 126€ to 145€ depending on the underlying price scenario. In the dynamic tariff case, maximum tolerable costs are lower and range from 50€ to 111€.

# Nomenclature

| | |
|---|---|
| $t$ | time (h) |
| $\Delta t$ | Time step used in calculations (h) |
| $\mathbb{T}$ | Set of all time steps considered |
| $\mathbb{T}_{out}$ | Set of time steps where outliers in price time series occur |
| $c_{2019}$ | Electricity price time series for the year 2019 (€ct/kWh) |
| $c_{out,pos}$ | Positive outliers of the electricity price time series (€ct/kWh) |
| $c_{scen}$ | Electricity price time series for the specified scenario (€ct/kWh) |
| $C_{scen,min}$ | Minimum price level of a given electricity price scenario (€ct/kWh) |
| $C_{scen,max}$ | Maximum price level of a given electricity price scenario (€ct/kWh) |
| $c_{adj}$ | Adjusted price time series without outliers (€ct/kWh) |
| $c_{out,adj,pos}$ | Adjusted price values for positive outliers (€ct/kWh) |
| $E_{grid,building}$ | Energy drawn from the grid (kWh) |
| $E_{PV,grid}$ | Energy fed into the grid from the PV system (kWh) |
| $E_H$ | The household's inflexible energy demand (kWh) |
| $E_{PV,building}$ | Energy supplied to the building by the PV system (kWh) |
| $E_{BSS}$ | Energy supplied to the building by the BSS (kWh) |

| Symbol | Description |
|---|---|
| $E_{EV}$ | Energy demand of the EV charging process (kWh) |
| $E_{HP,el}$ | Electrical energy demand of the heating system (kWh) |
| $\alpha_{PV}$ | Binary variable to indicate, if a PV system is available in the building |
| $\alpha_{BSS}$ | Binary variable to indicate if a BSS is available in the building |
| $\alpha_{EV}$ | Binary variable to indicate if an EV is available in the building |
| $\alpha_{HP}$ | Binary variable to indicate if a HP is available in the building |
| $c_{electricity\ price}$ | Electricity unit rate (€ct/kWh) |
| $c_{feed-in}$ | Feed-in remuneration (€ct/kWh) |
| $P_H$ | Power flow from the building to the household appliances (kW) |
| $P_{PV,grid}$ | Power flow from the PV system to the grid (kW) |
| $P_{PV,building}$ | Power flow from the PV system to the building (kW) |
| $P_{PV,BSS}$ | Power flow from the PV system to the battery storage system (kW) |
| $P_{PV,generation}$ | Power output of the PV system (kW) |
| $E_{BSS}$ | Energy stored in the battery storage (kWh) |
| $P_{BSS}$ | Power flow from the battery storage system to the building (kWh) |
| $\eta_{BSS,charge}$ | Charging efficiency of the battery |
| $\eta_{BSS,discharge}$ | Discharging efficiency of the battery |
| $q_{losses,BSS}$ | Standby losses of the battery storage system (%/h) |
| $P_{BSS,max}$ | Maximum charge/discharge power of the battery storage system (kW) |
| $E_{BSS,max}$ | Capacity of the battery storage system (kWh) |
| $E_{EV}$ | Energy stored in the EV's battery (kWh) |
| $P_{EV}$ | Charging power (kW) |
| $E_{EV,demand}$ | Energy demand of the EV (kWh) |
| $\eta_{EV,charge}$ | Charging efficiency of the EV's battery |
| $q_{losses,EV}$ | Standby losses of the EV's battery (%/h) |
| $E_{min,departure}$ | Minimum energy stored in the EV's battery at time of departure (kWh) |
| $f_{avail.}$ | Availability of the EV at the home location (binary) |
| $E_{min,charge}$ | Minimum energy level of the EV's battery, before charging process starts (kWh) |
| $E_{EV,max}$ | Capacity of the EV's battery (kWh) |
| $\dot{Q}_{heat\ demand}$ | Building's thermal heat demand (kW$_{th}$) |

| Symbol | Description |
|---|---|
| $\dot{Q}_{\text{HP,building}}$ | Thermal power flow from the heat pump to the building (kW$_{\text{th}}$) |
| $\dot{Q}_{\text{HS,building}}$ | Thermal power flow from the heat storage to the building (kW$_{\text{th}}$) |
| $COP$ | Coefficient of performance of the heat pump (p.u.) |
| $\eta_{\text{COP}}$ | Quality grade/scale-down factor of the heat pump's Carnot efficiency |
| $T_{\text{high}}$ | Flow temperature of the heating system (K) |
| $T_{\text{high,max}}$ | Technically maximum possible flow temperature of the heating system (K) |
| $T_{\text{amb.}}$ | Ambient temperature (K) |
| $T_{\text{amb.,norm}}$ | Norm outside temperature (K) |
| $T_{\text{room}}$ | Inside temperature (K) |
| $T_{\text{icing}}$ | Temperature threshold below which icing occurs (°C) |
| $f_{\text{icing}}$ | Factor of COP reduction due to icing (p.u.) |
| $\dot{Q}_{\text{max,th}}$ | Maximum thermal power output of the heat pump (kW$_{\text{th}}$) |
| $COP_{\text{nom}}$ | Nominal coefficient of performance (p.u.) |
| $\dot{Q}_{\text{nom,th}}$ | Nominal thermal power output of the heat pump (kW$_{\text{th}}$) |
| $P_{\text{HP,el}}$ | Power flow from the building to the heat pump (kW) |
| $\dot{Q}_{\text{HP,th}}$ | Thermal power generation of the heat pump (kW$_{\text{th}}$) |
| $E_{\text{HS}}$ | Energy stored in the heat storage (kWh$_{\text{th}}$) |
| $E_{\text{HS,max}}$ | Storage capacity of the heat storage (kWh$_{\text{th}}$) |
| $q_{\text{losses,HS}}$ | Standby losses of the heat storage (%/h) |
| $\dot{Q}_{\text{HP,HS}}$ | Thermal power flow from the heat pump to the heat storage (kW$_{\text{th}}$) |

# 1 Introduction

Amidst rising electricity prices[1] and the increasing volatility observed in day-ahead spot market prices in Germany over the past months (see Figure 1), households are becoming increasingly conscious of their energy expenditures.

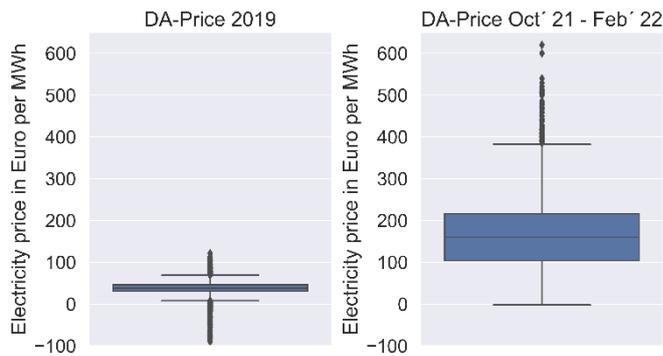

*Figure 1: Historic day-ahead spot market electricity prices for the year 2019 and October 2021 to February 2022*

Moreover, the increasing number of households adopting sector coupling technologies, such as electric vehicles (EVs) and heat pumps (HPs), lead to an increase in electricity consumption, prompting these households to seek ways to mitigate their overall energy expenditure. Additionally, higher electricity consumption and peak loads due to these technologies can pose challenges to local energy infrastructure. Managing high electricity costs and alleviating grid stress can be achieved through the flexible operation of EVs and HPs. The most prominent option discussed for incentivizing household customers to utilize their flexibility is dynamic electricity tariffs. Dynamic electricity tariffs are one option of a wide range of demand response (DR) mechanisms. DR mechanisms describe mechanisms, that lead to shifting or shedding of electricity demand to provide flexibility to help balance the grid. DR mechanisms can be beneficial for both end-users and utilities as they can increase overall system security and maximize social welfare [2]. DR mechanisms can generally be divided into incentive- or event-driven and price-driven mechanisms [3]. Incentive- or event-driven mechanisms include direct load control, emergency demand response programs, capacity market programs, interruptible/curtailable service, demand bidding/buyback programs, and ancillary service market programs [3]. The group of price-driven DR mechanisms includes all types of dynamic tariffs. One advantage of DR mechanisms is that they are implemented "behind the meter", on the customer side [4] and, in combination with smart meters, enable end-consumers to make informed decisions on their electricity consumption. To allow for an automated response to dynamic electricity tariffs, enabling technologies are necessary. These enabling technologies can generally be grouped into three categories: control devices, monitoring systems, and communication systems [5]. Control devices and monitoring systems are grouped into the home energy management system (HEMS). Together with smart meters as communication systems and dynamic electricity prices, these constitute the basic requirement for household customers to operate their flexible assets in a smart cost-optimized manner.

---

[1] Average price on the day-ahead spot market for Germany: 96,85€/MWh in the year 2021 and 235,45€/MWh in the year 2022 [1].

In the literature analyzing the interplay between DR mechanisms and dynamic electricity tariffs, two broad strands can be distinguished that address the following questions:

- What are the effects of a smart operation on households with regard to dynamic electricity prices and self-consumption?
- Which role do enabling technologies such as HEMS and smart meters play in the utilization of flexibility?

*Effects of a smart operation on households*

Lazar and Gonzalez [6] estimate a 54% reduction in peak load for households without any flexible technologies when dynamic pricing schemes are applied. Faruqui and Sergici [7] analyze various dynamic tariff pilot studies in different countries and find a peak load reduction of 3-6% for time-of-use (ToU) rates and 13-20% for critical peak pricing (CPP). In a study on a day-ahead real-time pricing (DA-RTP) scheme in Canada, Doostizadeh and Ghasemi [8] observe lower losses, a lower peak-to-peak distance, a flatter demand curve, and a higher load factor for household demand curves when compared to a flat tariff. When a PV system and BSS are combined with a ToU tariff that varies for different types of days and on energy drawn from the grid, Bignucolo et al. [9] find that applying the enhanced ToU tariff to both the power drawn from and fed into the grid can reduce peak load and daily electricity costs. A technology that has been discussed as very promising for increasing the utilization and financial benefits of dynamic tariffs is electric vehicles. With the principle of smart charging, Martinenas et al. [10] find that lower charging costs can be achieved when using a dynamic tariff. Aguilar-Dominguez et al. [11] investigate different dynamic tariff schemes (ToU, time-of-day, RTP) for households with an EV, a PV system, and a BSS. Results show a reduction in electricity costs of up to 85%, but an increase in peak load for all dynamic tariffs considered. Bonin et al. [12] study the combination of households with an EV, a PV system, and a DA-RTP tariff and define four charging strategies: direct charging, optimizing self-sufficiency, and optimizing electricity costs with the DA-RTP tariff without and with a PV system. The best result for electricity cost reduction is obtained for the optimizing self-sufficiency case (35% cheaper than direct charging), followed by the optimizing electricity costs with the DA-RTP tariff with a PV system case (34.5% cheaper). In both cases, there is an increase in peak load, which is lower with a PV system present. A more comprehensive take on the change in electricity costs is presented by Kühnbach et al. [13]. The study shows a change in electricity costs for all residential consumers by up to -3.7% when EVs in Germany apply controlled charging with a price signal derived from the national residual load. Included in the consideration is the change in electricity procurement costs due to controlled charging and the change in grid charges due to the necessary grid expansion with the uptake of EVs. According to Pena-Bello et al. [14], households with a HP and a PV system can increase their self-consumption rate and reduce the levelized cost of meeting the electricity demand between 13-26% when heat storage is added to the system. Ali et al. [15] show that energy cost savings are possible when a DA-RTP is used with electric space heating and partial thermal storage. Klaassen et al. [16] assume a DA-RTP tariff for a household with a HP for floor heating and DHW demand and apply a control algorithm to minimize heating costs. The results show an 8% reduction in energy heating costs but an increased heating demand of 1.4% (on average higher temperature in the HP system).

*Role of enabling technologies*

To effectively use dynamic tariffs and generate cost savings for household customers, automation in form of a HEMS including monitoring systems and control devices is necessary. A survey described by Buryk et al. [17] indicates that automated load shifting may play a critical role in encouraging customers to switch from a fixed to a dynamic tariff scheme. Automation can simplify the process for households, reducing the complexity and effort required to respond to dynamic tariffs, thus preventing response fatigue and increasing participation [18–21].

*Research gap and contribution*

The literature draws a clear picture regarding the general effects and direction of dynamic pricing: It reduces procurement costs and is far more effective when a smart and automated HEMS is used. However, as an increasing number of households is equipped with flexible technologies, we identify several gaps in the literature that need to be closed to evaluate the benefits of dynamic electricity tariffs more holistically:

Most studies focus on only one flexible technology, EV or HP (in combination with PV and or PV-BSS), and do not account for the interaction of both in one household when reacting to dynamic electricity prices. Furthermore, in most studies, only a few load curves are considered, while it is crucial to consider the heterogeneity of households to ensure a comprehensive understanding and generalizability of the impact of dynamic electricity prices on households with flexible appliances. Very often, HEMS and smart meters are taken as given in the studies. However, dynamic electricity prices and HEMS will only be adopted if the necessary investments do not exceed the cost savings earned through DR. Therefore, it is vital to also consider the associated costs of HEMS and smart meters. It is also crucial to not only compare the smart operation of flexible technologies responding to dynamic electricity prices with the case of no flexibility use at all but to also compare it to the case of smart operation of flexible technologies for self-consumption.

The outlined research and recent developments in electricity markets and digitization raise the following research questions:

- Considering multiple flexible technologies separately and jointly and accounting for different underlying average electricity prices and price spreads, do dynamic electricity prices for households have economic benefits?
- How does the utilization of flexibility in households compare between dynamic electricity tariffs and self-consumption as the current status quo for households with a PV system?
- Do possible cost savings compensate for the additional costs of metering point operation and HEMS?

To answer these questions, we developed the model *EVaTar-building*, a simulation model with embedded optimization, to map the household's electricity consumption when using a HEMS or reacting to dynamic tariffs. We consider two options: households using only a HEMS to increase self-consumption through a PV system or a PV-BSS, and households using a HEMS and a smart meter in combination with a dynamic electricity tariff based on the day-ahead spot market price. The dynamic tariffs under consideration are derived from the day-ahead spot market price of 2019. Additionally, we adjust the 2019 price time series to represent the prices recently seen on the day-ahead spot market in Germany, i.e. to produce higher average prices with higher price spreads. The model output allows us to analyze the economic effects, considering the investment in the HEMS and metering point operation costs in the case of using a smart meter. We also look at the changes in each household's annual electricity consumption and the corresponding load curves to get a comprehensive picture.

The study is organized as follows: Section 2 describes the methodological approach, the data used, and the underlying assumptions of the case study and the electricity price scenarios. Section 3 provides an overview of the study results. Section 4 discusses the results, and the study closes with the summary and conclusion (Section 5).

## 2   Material and Methods

We model the flexible and inflexible operation of different household types and their flexible technologies (e.g. EVs, HPs, PV-BSSs) with a HEMS based on a dynamic electricity tariff and a static feed-in tariff. We present the developed method for creating electricity price signals for purchasing electricity. Furthermore, we explain the calculation of the market value to represent the PV feed-in tariff. Both price signals are necessary for the analysis and allow for easy comparison of results across different scenarios. Finally, we provide a description of the model *EVaTar-building* and the characteristics of the household appliances and time series data that was used for this study.

### 2.1   Price signals

When it comes to using a HEMS, two price signals are relevant for household customers: the cost of electricity they are charged for each kWh drawn from the grid, and the feed-in tariff they receive for each kWh their PV system feeds into the grid[2].

---

[2] This assumes the current regulatory framework in Germany, where electricity is paid in €ct/kWh without any costs for the capacity used

### 2.1.1 Electricity price time series manipulation and scenarios

#### 2.1.1.1 Electricity price time series manipulation

To make our results more generalizable, our goal is to look at dynamic electricity tariffs using different underlying price scenarios. To allow a comparison between these scenarios, we pick one year, in our case 2019[3] (corresponding roughly to the prices during the last decade), and manipulate the price time series while preserving its underlying structure. Since there are some hours in the year with comparably high and low prices, we define the 1% percentile of all electricity prices in the time series as outliers, which gives us a set of time steps $\mathbb{T}_{\text{out}}$ in which these outliers occur. Subsequently, we adjust the mean value and standard deviation of the remaining dataset for the chosen scenarios according to Eq. (1).

$$c_{\text{adj}}^t = \frac{c_{2019}^t - \overline{c_{2019}^t}}{\sigma(c_{2019}^t)} \cdot \sigma_{\text{scen}} + \emptyset_{\text{scen}}, \qquad \forall t \notin \mathbb{T}_{\text{out}} \tag{1}$$

with $c_{2019}^t$ being the electricity price of the given time series for each time step $t$ excluding the timesteps where outliers occur, $\sigma_{\text{scen}}$ the standard deviation and $\emptyset_{\text{scen}}$ the mean value of the electricity price for the specified scenario. $c_{\text{adj}}^t$ represents the adjusted time series (without outliers).

In a next step, we adjust the outliers to lie between the given minimum and maximum values of the price scenario ($C_{\text{scen,min}}$ and $C_{\text{scen,max}}$) and the minimum and maximum values of the adjusted time series ($\min c_{\text{adj}}^t$ and $\max c_{\text{adj}}^t$). This is done for positive and negative values separately. Eq. (2) is exemplary for positive values.

$$c_{\text{out,adj,pos}}^t =$$
$$\begin{cases} \frac{c_{\text{out,pos}}^t - \min c_{\text{out,pos}}^t}{\max c_{\text{out,pos}}^t - \min c_{\text{out,pos}}^t} \cdot \left(C_{\text{scen,max}} - \max c_{\text{adj}}^t\right) + \max c_{\text{adj}}^t, \forall c_{\text{out,pos}}^t \geq c_{\text{adj}}^t, \forall t \in \mathbb{T}_{\text{out}} \\ \max c_{\text{adj}}^t, \forall c_{\text{out,pos}}^t < c_{\text{adj}}^t, \forall t \in \mathbb{T}_{\text{out}} \end{cases} \tag{2}$$

After completing the manipulation of the separated datasets $c_{\text{adj}}^t$, $c_{\text{out,adj,pos}}^t$ and $c_{\text{out,adj,neg}}^t$, they are all put back together to one time series $c_{\text{scen}}^t$ representing the electricity price scenario.

---

[3] 2019 is chosen to allow for a consistent scenario also in combination with available data for weather, temperature etc. for the flexible consumers, as comprehensive data for more recent years was not available to the authors.

### 2.1.1.2 Electricity price scenarios

We define three scenarios:

- a **low price scenario** that corresponds to the 2019 time series,

- a **high price scenario** that uses values from October 2021 to February 2022[4] (see Figure 1),

- and a **medium price scenario** that lies between these two.

The mean values, standard deviations, and minimal and maximum values for the resulting time series after the electricity price time series manipulation described in Section 2.1.1.1 are listed in Table 1.

Table 1: Mean values, standard deviation, and minimal and maximal values of the electricity price scenarios used

|  | Mean | Standard deviation | Minimal value | Maximal value |
|---:|---|---|---|---|
| Low price scenario | 22.6 €ct/kWh | 1.8 €ct/kWh | 7.4 €ct/kWh | 32.6 €ct/kWh |
| Medium price scenario | 30.2 €ct/kWh | 6.2 €ct/kWh | 4.4 €ct/kWh | 48.3 €ct/kWh |
| High price scenario | 37.8 €ct/kWh | 10.7 €ct/kWh | 1.5 €ct/kWh | 91.9 €ct/kWh |

Figure 2 (a) shows average prices for each hour of the day seen by a household (including taxes, levies, and surcharges), broken down by season and the range from minimum to maximum value for all three price scenarios. The prices are seasonally dependent, particularly in the range between the minimum and maximum values. For all seasons the lowest prices can be seen in the early morning hours and – to some extent – in the early afternoon. Figure 2 (b) shows the intraday and intraweek price spreads, which exhibit a clear seasonal difference in the intraweek spreads, with smaller differences in the intraday spreads.

---

[4] Values from October 2021 to February 2022 are used, as data for a time period beyond February 2022 was not available at the point of the creation of this study and the sharp increase in spot market prices was seen approx. from October 2021.

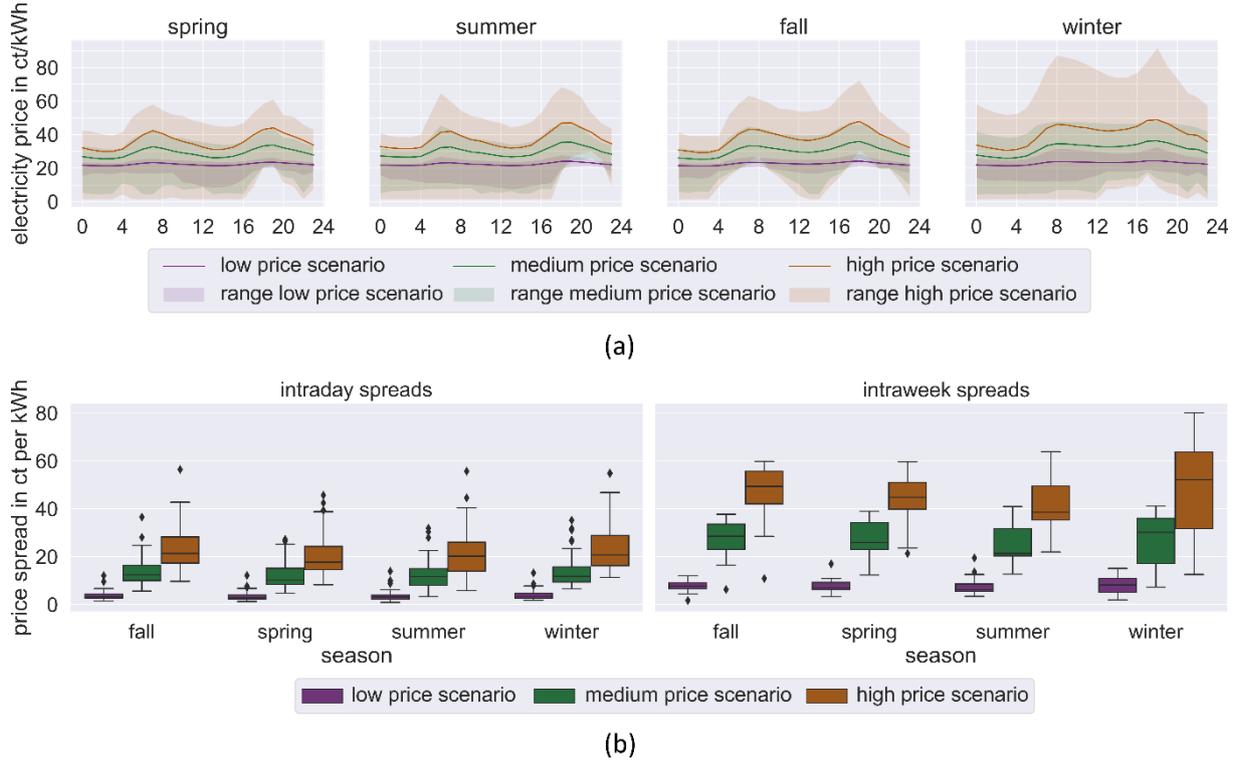

*Figure 2: (a) Average prices and range for each hour of the day broken down by season for the three price scenarios; (b) intraday and intraweek price spreads for the three price scenarios*

To provide a basis for comparison, a static tariff for each price scenario is defined, which corresponds to the level of the mean value of the dynamic tariff (see Table 1). In the context of existing electricity pricing structures in Germany, static tariffs exhibit a higher average value compared to dynamic tariffs. However, utilizing the mean value of dynamic tariffs as proxy for static tariffs is deemed justifiable. This approach serves as a worst-case scenario from the standpoint of flexibility utilization.

### 2.1.2 PV feed-in tariff

With changing prices at the day ahead spot market, the market value of electricity generated from PV systems will adapt as well. To account for that in the different scenarios, we consider a constant feed-in remuneration for the generated power from PV rooftop systems. To allow for easy comparison between the scenarios, we calculate a PV market value $MV_{\text{solar}}$ based on the day ahead spot market prices. The electricity price $c^t_{\text{electricity}}$ is weighted by the amount of electricity generated by PV systems $E^t_{\text{PV}}$ at each hour $t$ (Eq. (3)).

$$MV_{\text{solar}} = \frac{\sum_{t=0}^{8760} E^t_{\text{PV}} \cdot c^t_{\text{electricity}}}{\sum_{t=0}^{8760} E^t_{\text{PV}}} \qquad (3)$$

The feed-in from PV systems will be remunerated with the market value of the underlying price scenario (see Table 2)[5].

Table 2: PV market values for the three electricity price scenarios

|  | PV market value $MV_{solar}$ |
|---|---|
| Low price scenario | 3.62 €ct/kWh |
| Medium price scenario | 9.71 €ct/kWh |
| High price scenario | 15.81 €ct/kWh |

## 2.2 Flexible consumer model *EVaTar-building*

Our computational model incorporates various components: inflexible household demand, battery storage operation, and demand response capabilities of electric vehicles and heating systems. The home energy management system is formulated as a mixed-integer linear program aimed at minimizing electricity costs.

In the model, each building represents a single household and may feature additional technologies such as EVs, PV systems, battery storage systems, and heating systems (HS), which consist of a heat pump and a heat storage tank. Connected to a low-voltage grid, each building can either draw electricity from the grid or feed electricity generated by the PV system back into it. Figure 3 outlines the fundamental architecture and energy flows of a flexible consumer within the *EVaTar-building* model.

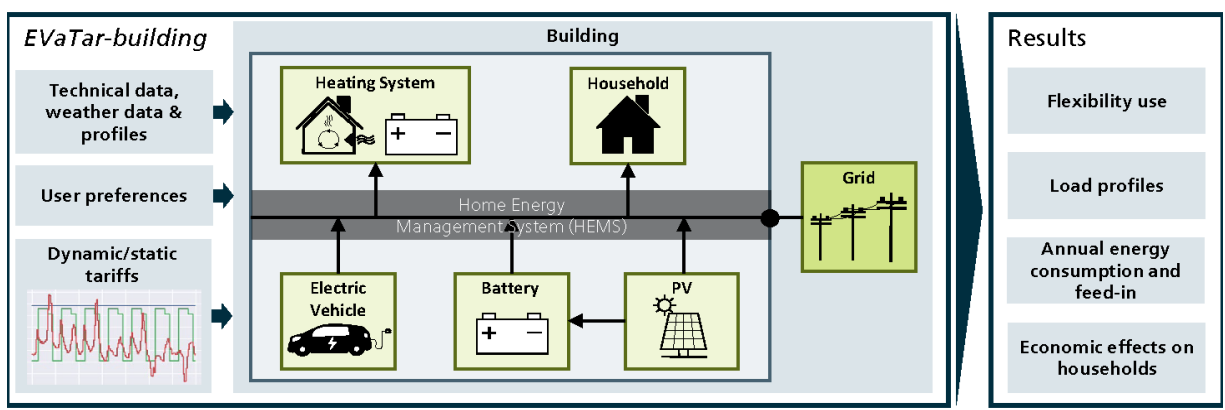

Figure 3: Schematic overview of the flexible consumer model EVaTar-building

We operate under the assumption that each household can be equipped with a HEMS that functions with perfect foresight. The HEMS actively engages in building operations with the primary goal of minimizing the electricity purchase costs. The corresponding objective function is articulated in Equation (4):

$$\min \sum_{t=0}^{t_{\max}} E_{\text{grid,building}}^t \cdot c_{\text{electricity price}}^t - E_{\text{PV,grid}}^t \cdot c_{\text{feed-in}} \quad (4)$$

---

[5] A constant value is selected because time-varying remuneration is generally not observed for smaller photovoltaic systems installed in single-family homes in Germany.

Here, $E_{\text{grid,building}}^t$ represents the amount of energy drawn from the grid during hour $t$. $c_{\text{electricity price}}^t$ denote the applicable electricity price during this time unit. $E_{\text{PV,grid}}^t$ signifies the energy fed back into the grid from the PV system during hour $t$. $c_{\text{feed-in}}$ is the time-independent feed-in remuneration. The energy drawn from the grid $E_{\text{grid,building}}^t$ is specified as follows:

$$E_{\text{grid,building}}^t = E_{\text{H}}^t - \alpha_{\text{PV}} \cdot E_{\text{PV,building}}^t - \alpha_{\text{BSS}} \cdot E_{\text{BSS}}^t + \alpha_{\text{EV}} \cdot E_{\text{EV}}^t + \alpha_{\text{HP}} \cdot E_{\text{HP,el}}^t, \\ \forall t \in \text{T} \quad (5)$$

In this context, $\alpha$ is a binary variable indicating whether a particular technology is present in a household. $E_{\text{H}}^t$ refers to the household's inflexible electricity demand during hour $t$. $E_{\text{PV,building}}^t$ signifies the amount of energy supplied to the building from the PV system during hour $t$. $E_{\text{BSS}}^t$ denotes the energy provided to the building by the battery storage system in hour $t$. $E_{\text{EV}}^t$ represents the energy requirements of the EV charging process during hour $t$. Finally, $E_{\text{HP,el}}^t$ specifies the electrical energy demand of the heating system during hour $t$.

To account for uncertainties in load and generation forecast, the model employs a rolling horizon scheme that considers a 3-day horizon with an optimization interval of 24 h.

The model encompasses a wide range of household customers, in particular the households themselves, their PV systems, PV-BSS, EVs, and heating systems. These components of the considered flexible consumers are explained in the following sections.

### 2.2.1 Household appliances (H)

In this study, the electricity demand of household appliances is considered inflexible and is defined exogenously in the form of household load profiles. Consequently, the households' appliances cannot be controlled by the HEMS.

To cover the heterogeneity of household load profiles, we use residential load profiles of 316 households from a smart meter field study carried out in Austria and Germany [22]. An overview of the annual electricity consumption of the considered households can be found in Figure Annex 1(a).

### 2.2.2 PV system (PV)

The installed PV system can be used within the model to supply the various applications in the building, charge a battery storage system, and feed electricity into the grid. To represent this, the power output from the PV system $P_{PV,generation}^t$ for a timestep $t$ is separated into three different power flows (Eq. (6)).

$$P_{PV,grid}^t + P_{PV,building}^t + P_{PV,BSS}^t = P_{PV,generation}^t, \quad \forall t \in T \tag{6}$$

Where $P_{PV,grid}^t$ is the power flow from the PV system to the grid, $P_{PV,building}^t$ represents the power flow from the PV system to the building and $P_{PV,BSS}^t$ is the power flow from the PV system to the battery storage system. Buildings consisting of only a household and a PV system, i.e. buildings without any of the flexibility options presented in the following, are considered inflexible.

For buildings with a PV system, we adjust the installed capacity to the annual electricity consumption of each household. We do this by using information from [23]. For an overview of the installed capacity for all PV systems considered, see Figure Annex 3(a). The PV generation profile is taken from renewables.ninja[6] for the year 2019 and the city of Karlsruhe.

### 2.2.3 Battery storage system (BSS)

The battery storage system of a building can be charged with electricity from the PV system. It can also supply power to the building's various applications.

The energy currently stored in the battery $E_{BSS}^t$ at an hour $t$ is defined as follows:

$$E_{BSS}^t = (1 - q_{losses,BSS}) \cdot E_{BSS}^{t-1} + P_{PV,BSS}^t \cdot \eta_{BSS,charge} \cdot \Delta t - P_{BSS}^t \cdot \frac{1}{\eta_{BSS,discharge}} \cdot \Delta t, \tag{7}$$

$$\forall t \in T$$

With:

---

[6] Renewables.ninja is an open-source tool that considers historical weather characteristics and technical parameters to calculate supply profiles. For PV, we use an azimuth angle of 180° and a tilt of 35°. Provided by [24] and described in [25] and [26].

$$0 \leq P_{PV,BSS}^t \leq P_{BSS,max}, \qquad \forall t \in T \qquad (8)$$

$$0 \leq P_{BSS}^t \leq P_{BSS,max}, \qquad \forall t \in T \qquad (9)$$

$$0 \leq E_{BSS}^t \leq E_{BSS,max}, \qquad \forall t \in T \qquad (10)$$

$P_{PV,BSS}^t$ is the power flow from the PV system to the battery storage system, $P_{BSS}^t$ is the power flow from the battery to the building, $P_{BSS,max}$ is the maximum charge/discharge power of the battery system, and $E_{BSS,max}$ is the maximum usable capacity of the battery storage system. $\eta_{BSS,charge}$ and $\eta_{BSS,discharge}$ are the charging and discharging efficiency factors and $q_{losses,BSS}$ depicts the standby losses.

Households with both a PV system and a BSS can operate using two different strategies:

1. The first, smart operating strategy allows for flexible and predictive operation with dynamic tariffs. In this strategy, the BSS is included in the building's HEMS and thus used to minimize the overall energy procurement costs of the building (see Eq. (4))[7].
2. The second strategy is currently the most common operating strategy for residential PV-BSS. The strategy is independent of the underlying tariff system and is used by households without a HEMS. The electricity generated by the PV system is first used to meet the electricity demand of the building, including the electricity demand of the EV and the heating system. If there is a generation surplus, it is used to charge the battery storage. Any further surplus is fed into the grid. When the building's electricity demand exceeds the PV generation, the electricity is drawn from the battery to supply the household's energy demand.

To determine the usable battery capacity of the battery for each household with a PV system, we consider two factors: the household's annual electricity consumption and the installed capacity of their PV system. The correlation between these factors and the usable battery capacity is obtained from [23]. The resulting distribution of usable battery capacity for the households considered can be found in Figure Annex 3(b). For all battery storage systems, we assume a C-rate of 1 and a round-trip efficiency of 0.95 and standby losses of 0.01% per hour.

---

[7] Battery-to-grid power supply is not included in our analysis. This exclusion is based on the fact that the battery can only be charged via the PV system, and given the constant feed-in remuneration, employing the battery would merely result in additional energy losses.

### 2.2.4 Electric vehicle (EV)

Electric vehicles are implemented as mobile battery storage, which means that the EV battery is not available as a flexibility resource at all times. Driving profiles with the corresponding energy demand $E_{EV,demand}^t$ are assigned to the studied households. The assignment is based on the socio-demographic metadata of the households and the driving profiles. The hourly availability of a vehicle at the home location $f_{avail.}^t$ is defined exogenously.

The energy currently stored in the EV battery $E_{EV}^t$ is defined as follows (Eq. (11)):

$$E_{EV}^t = (1 - q_{losses,EV}) \cdot E_{EV}^{t-1} + P_{EV}^t \cdot \Delta t \cdot \eta_{EV,charge} - E_{EV,demand}^t, \qquad \forall t \in T \qquad (11)$$

Here $\eta_{EV,charge}$ is the efficiency of the charging process and $P_{EV}^t$ depicts the charging power at a time step $t$, and $q_{losses,EV}$ stands for the standby losses of the EV battery.

There are two different charging strategies for EVs:

1. The first strategy involves controlled charging of the EVs, where the EV is integrated into the building's HEMS. EV owners can enter their preferences into the HEMS, such as the minimum range or minimum energy stored in the battery at the time of departure $E_{min,departure}$ (Eq. (12)) and the minimum energy level $E_{min,charge}$ at which they want to start charging their vehicle at the latest, (Eq. (13)). These preferences are represented in the optimization by the following constraints:

$$E_{EV}^t \geq E_{min,departure}, \qquad \forall t \in T \text{ where } f_{avail.}^{t-1} == 1 \wedge f_{avail.}^t == 0 \qquad (12)$$

$$E_{min,charge} \leq E_{EV}^t \leq E_{EV,max} \; \forall t \in T \qquad (13)$$

Where $E_{EV,max}$ represents the maximum capacity of the EV's battery.

2. The second strategy involves continuous charging of the EV immediately after arrival until a state of charge of 100% is reached or the EV starts driving again.

The availability at the home location and the power output while driving EVs are taken from Ref. [27]. The data is computed with the vehicle diffusion model "ALADIN"[8], which uses vehicle usage data from Ref. [31]. Figure Annex 2(c) gives an overview of the yearly mileage of all EVs considered. The energy consumption while driving ranges from 0.15 to 0.20 kWh/km and the usable battery capacity from ranges 34.2 to 90.0 kWh. $E_{\text{min,departure}}$ is set to 80% of the maximum range, $E_{\text{min,charge}}$ is set to 20% of the maximum range. The charging power of all EVs is set to 11 kW. Standby losses are assumed to be 0.01% per hour.

---
[8] For more information on the ALADIN model, we refer to [28], [29], and [30]

### 2.2.5 Heating system

The heating system of a building consists of an air-to-water heat pump (HP) and a heat storage tank (HS). The heat pump can supply energy to both the building ($\dot{Q}^t_{\text{HP,building}}$) and the heat storage tank ($\dot{Q}^t_{\text{HP,HS}}$). The heat storage tank can only supply the building ($\dot{Q}^t_{\text{HS,building}}$). The building's heat demand $\dot{Q}^t_{\text{heat demand}}$ must be met for each time step $t$ (Eq. (14)).

$$\dot{Q}^t_{\text{heat demand}} = \dot{Q}^t_{\text{HP,building}} + \dot{Q}^t_{\text{HS,building}} \tag{14}$$

The heating system can be used inflexibly or flexibly. In the latter case, the heating system is included in the HEMS.

*Heat pump (HP)*

The flow temperature of the heating system $T_{\text{high}}$ follows a heating curve and is, therefore, dependent on the ambient temperature. It is calculated as follows (Eq. (15)):

$$T^t_{\text{high}} = T_{\text{room}} + \left(T_{\text{high,max}} - T_{\text{room}}\right) \cdot \left(\frac{T_{\text{room}} - T^t_{\text{amb.}}}{T_{\text{room}} - T_{\text{amb.,norm}}}\right)^{1/n} \tag{15}$$

where $T_{\text{room}}$ is the inside temperature, which is considered to be held constant by the heat pump system, $T_{\text{high,max}}$ represents the technically maximum possible flow temperature of the heat pump, $T_{\text{amb.,norm}}$ is the norm outside temperature for the given location and $n$ is the radiator exponent which is set to $n = 1.33$ for wall mounted radiators.

Furthermore, a temperature-dependent coefficient of performance $COP^t$ is assumed for the air-to-water heat pump. In addition, we account for efficiency losses due to icing at ambient temperatures below 2°C ($T_{icing}$) by reducing the COP for low temperatures by a factor of $f_{\text{icing}} = 0.2$. With a typical quality grade for air-to-water heat pumps of $\eta_{\text{COP}} = 0.4$ [32], the COP is calculated as follows (Eq. (16)):

$$COP^t = \begin{cases} \eta_{COP} \cdot \dfrac{T_{high}}{T_{high} - T_{amb.}^t}, & T_{amb.}^t > T_{icing} \\ \eta_{COP} \cdot \dfrac{T_{high}}{T_{high} - T_{amb.}^t} \cdot (1 - f_{icing}), & T_{amb.}^t \leq T_{icing} \end{cases}, \quad \forall t \in T \quad (16)$$

where $T_{high}$ represents the flow temperature of the heating system and $T_{amb.}^t$ the ambient temperature at a given time step $t$.

With the time-dependent $COP^t$, the nominal $COP_{nom}$ and the nominal power output $\dot{Q}_{nom,th}$ of the heat pump, the maximum possible power output $\dot{Q}_{max,th}^t$ can be calculated for each time step:

$$\dot{Q}_{max,th}^t = \frac{COP^t}{COP_{nom}} \cdot \dot{Q}_{nom,th} \quad (17)$$

The necessary electric power consumption from the building to the heat pump $P_{HP,el}^t$ is defined as the ratio of the thermal power generation $\dot{Q}_{HP,th}^t$ of the heat pump and the COP:

$$P_{HP,el}^t = \frac{\dot{Q}_{HP,th}^t}{COP^t} \quad (18)$$

To determine the appropriate size of the heat pump for each building, we use the living space of the building, which was obtained from the smart meter field study mentioned in Section 3.1.1[9]. The heat pump is sized to match the heating demand of the building, which is assumed to be 100 kWh/m2/a. The heat demand profile is obtained from HotMaps [33] for the city of Karlsruhe (DE12) and is scaled based on the annual heat demand of each building. For consistency, the ambient temperature for the same year (2019) and location is obtained from the Climate Data Center of the German Weather Service (Deutscher Wetterdienst) for Station ID 4177 [34] and used to model the heating system. $T_{high,max}$ is set to 50°C and $\eta_{COP}$ is assumed to be 0.4.

*Heat storage (HS)*

The heat storage tank is used to make the heating system more flexible. Its storage capacity $E_{HS,max}^t$ is defined so that it can store the energy of two hours of the maximum power output of the heat pump[10].

---

[9] An overview on the living space of all households considered is given in Figure Annex 1(b).
[10] This corresponds to a typical design parameter for heat storage tanks in Germany. Heat storage tanks of this size are compact enough to be retrofittet into most single-familiy homes. At the same time, they allow for the bridging of restricted periods as per §14a of the German Energy Act (EnWG).

The energy currently stored $E_{HS}^t$ in the heat storage is defined as (Eq. (19)):

$$E_{HS}^t = (1 - q_{losses,HS}) \cdot E_{HS}^{t-1} + \dot{Q}_{HP,HS}^t \cdot \Delta t - \dot{Q}_{HS,building}^t \cdot \Delta t \tag{19}$$

Here $q_{losses,HS}$ represents the factor of the standby losses of the heat storage.

The heat storage is designed to store enough energy to meet the maximum heat demand of the building for two consecutive hours. Standby losses of the heat storage are assumed to be 0.2% per hour.

## 2.3 Case study

We examine three different cases:

- **No-flex case:** This is the base case, where all households have a static tariff and do not utilize their flexibility.
- **Self-consumption flex (SC-flex) case:** Here, households use a HEMS to reduce electricity costs by increasing their self-consumption from a PV- or PV-BSS-system. Households do still use the static tariff.
- **Dynamic tariff flex (DT-flex) case:** In this case, households use a HEMS to reduce electricity costs. They use a dynamic tariff and therefore are additionally equipped with a smart meter.

Each case is simulated for each electricity price scenario described in Section 2.1, resulting in nine observations. The technology combinations considered are households with EV, HP, or EV and HP, each without further technology, with a PV system or a PV-BSS. For each observation, results are analyzed for each of the 316 households and nine technology combinations.

In the first step, we have a look at the effects on the households' load curves to see how (e.g. from/to which hours) load is shifted by the HEMS with and without a dynamic electricity tariff. Secondly, we analyze how the electricity drawn from the grid and the self-consumption rates change with the different electricity price scenarios and cases. From there, we show the effects on the household's annual unit rate costs and extend the analysis to include not only unit rate costs but also investment costs for the HEMS and metering point operation costs of smart meters. In the last step, we examine the maximum tolerable costs for enabling technologies (HEMS, smart meter), so that 75% of all households considered would still see financial benefits from utilizing their flexibility, either in the SC-flex or the DT-flex case.

# 3 Results

## 3.1 Effects on load curves

The first thing to look at is the change in the load curves of households when flexibility is utilized. The load curves for the SC-flex and the DR-flex case are presented in Figure 4.

The hourly averaged load curves of all households with EVs show a clear shift from the evening hours to the early morning hours for the DT-flex case due to the lowest electricity prices occurring in those hours. The effect is slightly reduced when households have an additional PV system or a PV-BSS, as self-generated electricity can be used for charging the EV. The load curves do not differ noticeably between the different price scenarios. For the SC-flex case, we observe a load shift from the evening to the morning hours. EVs are charged later than in the no-flex case, due to the HEMS also considering self-discharge of vehicle batteries. This observation can be made for hours with and without PV generation.

For households with a HP, a load shift to the early morning hours (lowest electricity prices) and the afternoon is observed for the DT-flex case. The interplay of electricity prices and temperature-dependent COP and heating curve plays a decisive role here as the HP can be operated more efficiently during the day when the outside temperature is higher and, therefore, the COP is higher. This COP effect is more pronounced in the low price scenario than in the medium or high price scenario as price spreads are lower. This means, that the higher the price spreads/peaks, the more likely it is that a lower COP is accepted. Adding a PV system or a PV-BSS to households with a HP will lead to self-generated electricity being used to power the heating system. Nevertheless, we still see a slight increase in load in the early morning hours, which shows a reaction to the dynamic tariff. The effects for households with a HP are higher in winter, where outside temperatures are lowest and therefore heating demand is highest. See Figure Annex 1 for a detailed evaluation by season for each technology combination.

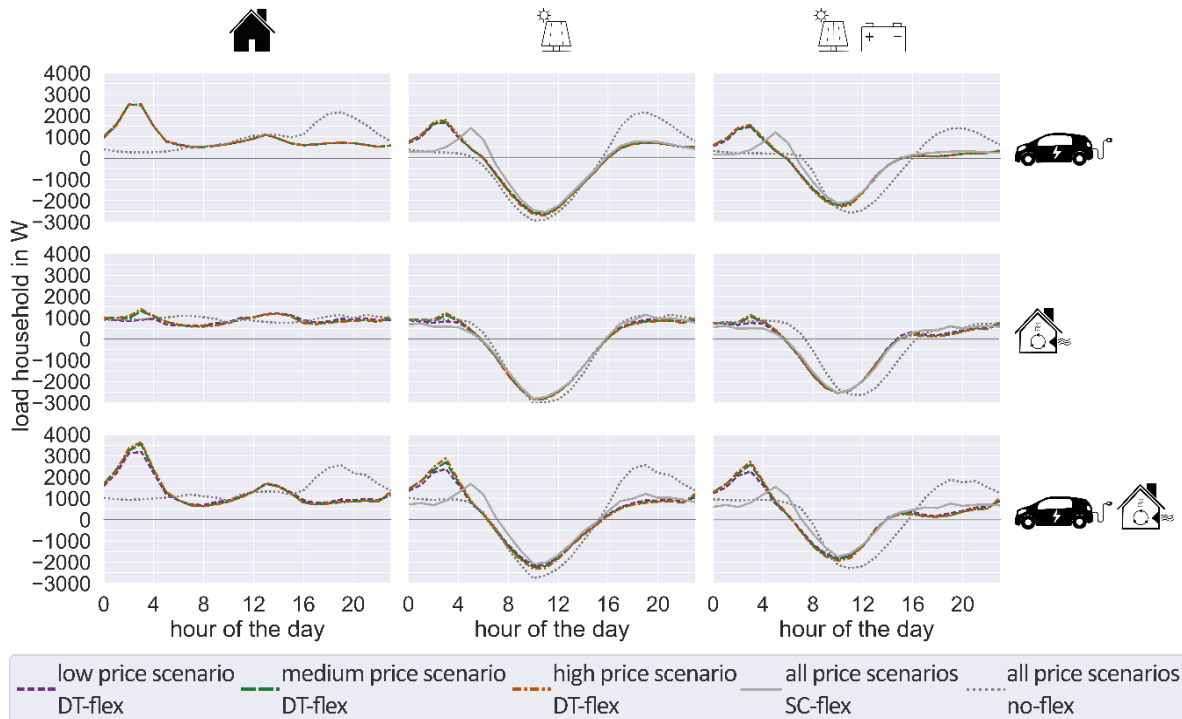

Figure 4: Yearly average household load for each hour of the day including all households considered for the defined cases and scenarios.

## 3.2 Electricity drawn from the grid and self-consumption

The load curve analysis clearly shows that the utilization of flexibility by HEMS results in more efficient utilization of HPs. This is reflected in Figure 5(a), which presents the change in annual electricity drawn from the grid for the DT-flex case across all three price scenarios. For households with only an EV, there are no significant changes in electricity drawn from the grid, as the charging process of EVs is assumed to be equally efficient at all hours of the year. However, for households with a HP, there is a clear reduction of up to 10.9% in annual electricity drawn from the grid, which shows the more efficient utilization of hours with higher COP values to power the heating system. For households with an additional PV system or a PV-BSS, an increase in the self-consumption rate can be achieved for both the SC-flex and the DT-flex case in all price scenarios (Figure 5). Self-consumption rates of up to 62.1% (for households with EV, HP, and PV-BSS) can be achieved. In the low price scenario, there are no notable differences in self-consumption rate between the SC-flex and the DT-flex case for all technology combinations. However, this changes with higher price scenarios. As seen before, the higher price scenarios lead to the HEMS shifting more load to hours with lower electricity prices instead of only increasing self-consumption. Consequently, in the high price scenario, we see a lower self-consumption rate for the DT-flex case for all technology combinations.

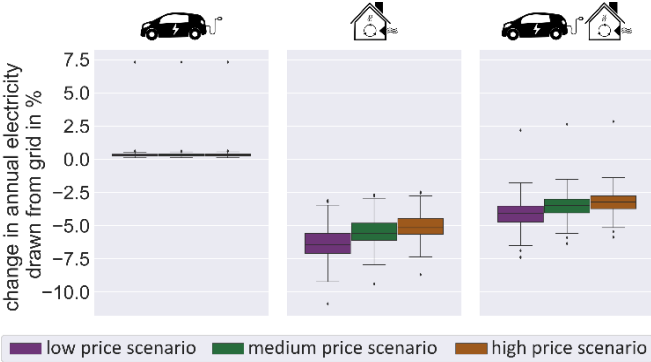

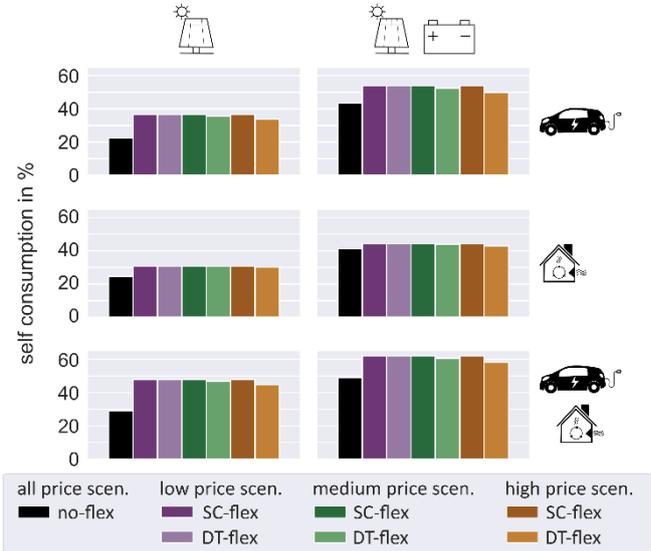

*Figure 5: (a) Change in annual electricity drawn from the grid for the DT case for all three price scenarios; (b) Average self-consumption rate for the SC-flex and DT-flex case for all three price scenarios*

## 3.3 Annual unit rate costs

The households' annual unit rate costs are affected by the changes in self-consumption, annual electricity drawn from the grid, and the use of dynamic tariffs. The SC-flex case shows a cost reduction in all price scenarios compared to the no-flex case (Figure 6). Households with an EV benefit more than households with a HP. By adding the dynamic tariff in the DT-flex case, further cost reductions can be achieved in almost all cases, with households with an EV benefiting more. However, households with a HP and a PV system on average see an increase in annual unit rate costs in the DT-flex case compared to the SC-flex case. This is due to the HEMS not being able/allowed to shift the load of HPs as freely as the load of EVs because HPs must follow a relatively steady heat demand of the household with a comparably small storage capacity. Consequently, electricity consumption cannot completely be shifted away from hours with high prices (especially in the evening hours). This leads to higher overall costs, as the unit rate of the static tariff in those hours is lower than the unit rate of the dynamic tariff.

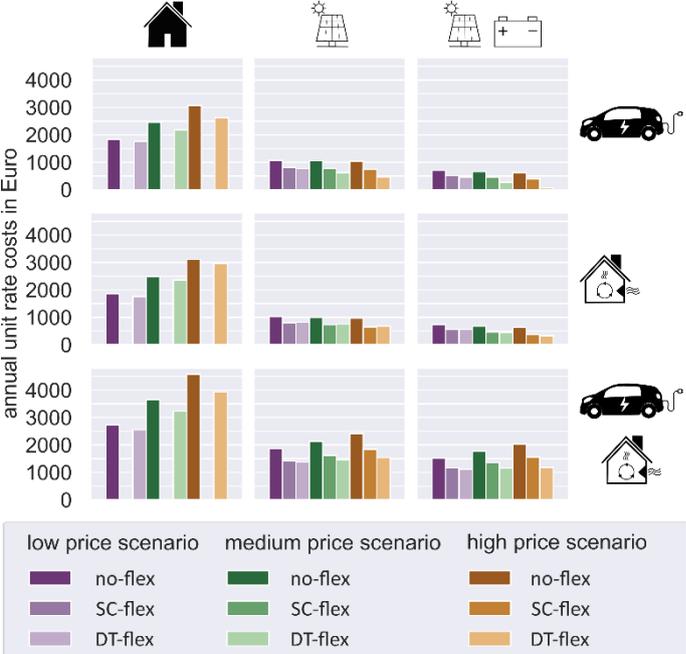

*Figure 6: Annual unit rate costs for the no-flex, SC-flex, and DT-flex cases for all three price scenarios*

## 3.4 Financially most attractive options – incl. costs for HEMS and smart meter

To fully evaluate the financial effects of the SC-flex and DT-flex cases with the given price scenarios, it is important to consider not only the annual unit rate costs but also additional costs such as the investment in the HEMS for the SC-flex and DT-flex case and the metering point operation costs for the smart meter for the DT-flex case. By allocating these costs to the annual unit rate costs, the most financially attractive option for each household can be determined in all price scenarios. Assuming an investment of 1,500€ for the HEMS, yearly annualized costs of around 167 €/yr are expected[11]. Adding 7.5 €/month for metering point operation leads to annual costs of 90€. Figure 7 shows the share of households opting for each case in each price scenario.

Higher price scenarios increase the financial attractiveness of using a HEMS and dynamic tariff, with households having an EV benefiting more from dynamic tariffs than households with a HP. The potential of self-consumption optimization with a HEMS is greater for households with a HP. In the high price scenario, 85.4% of households with an EV opt for the DT-flex case, while only one household with a HP would choose this option. The addition of a PV system makes the SC-flex case the most financially attractive option for all households with a HP in the medium and high price scenarios. For households with both an EV and a HP, using dynamic tariffs yields even greater benefits. In the medium price scenario, 93.7% of these households see the DT-flex case as the financially most attractive option, rising to 99.4% in the high price scenario. Finally, adding a PV system or a PV-BSS to households with an EV, a HP or both an EV and a HP can further increase the financial attractiveness in the SC-flex and the DT-flex case.

---

[11] With an assumed lifetime of 10 years and an interest rate of 2%

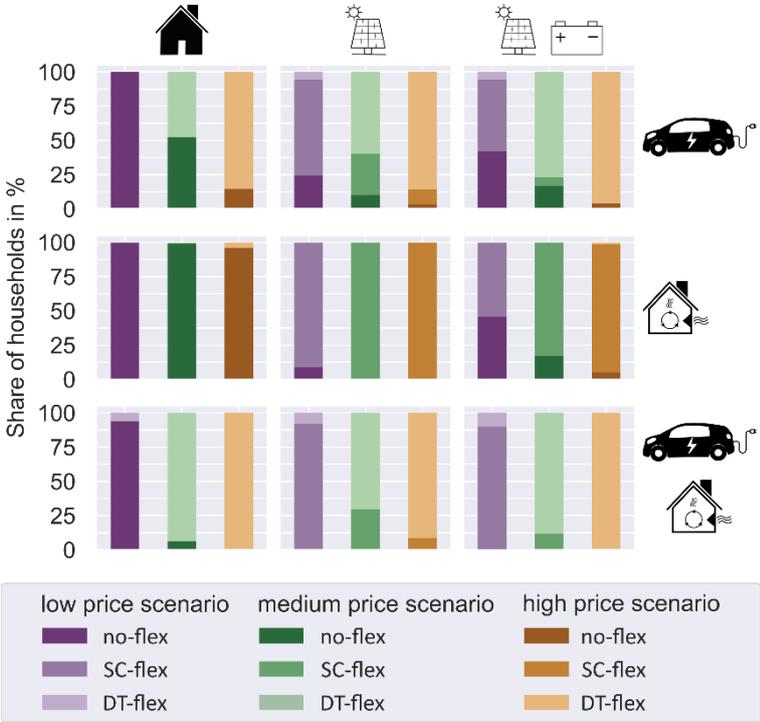

*Figure 7: Share of households having the three cases considered as the financially most attractive option for all three price scenarios.*

## 3.5 Maximum tolerable costs of HEMS and smart meter for incentivizing flexibility

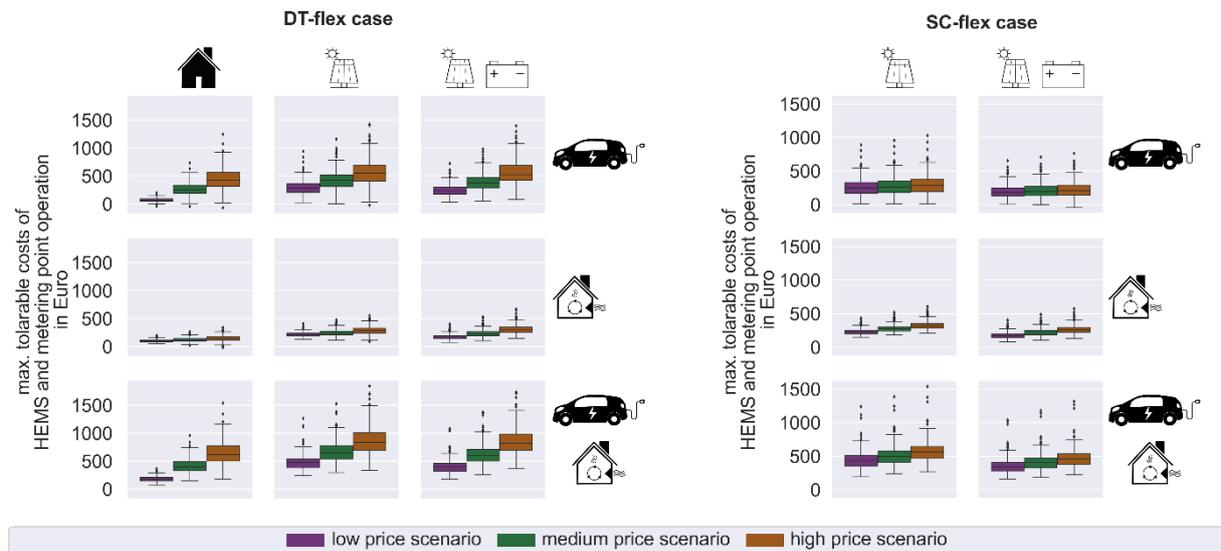

*Figure 8: Max. tolerable costs of HEMS and metering point operation to make the utilization of flexibility financially attractive for households. Results are shown for the DT-flex (left) and the SC-flex (right) cases for all three price scenarios.*

The analysis shows that the investment costs in a HEMS and metering point operation costs are significant factors in determining the financial attractiveness of using a HEMS but also dynamic tariffs. The question arises, of how high the costs may be to still incentivize their use. To answer this, we assume that households will invest in a HEMS or a HEMS in combination with dynamic tariffs in the SC-flex and DT-flex case when they can cover the costs of HEMS and metering point operation through savings in their annual unit rate costs compared to the no-flex case. Therefore, we define the maximum tolerable costs of the HEMS and metering point operation as the annual unit rate cost savings.

By determining the maximum tolerable costs for all technology combinations and all three price scenarios, it can be seen that the maximum tolerable costs vary greatly between the individual households (min/max values of each boxplot), with less variance for households with a HP than for households with an EV (Figure 8). As in reality, it will be unlikely to incentivize all households to utilize their flexibility, we focus on the 75% quantile of all households in the following analysis (see Table 3).

In the SC-flex case, households with an EV and a PV system can tolerate maximum annual costs for the HEMS of around 171€ (low price scenario) to 196€ (high price scenario), which get reduced to 126€ to 145€ if a BSS is added. For households with a HP, a PV system or PV-BSS maximum tolerable costs for the HEMS are around 146€ (HP and PV-BSS, low price scenario) to 282€ (HP and PV system, high price scenario). Households with both an EV and a HP allow for even higher maximum tolerable costs for the HEMS. In the DT-flex case, where metering point operation costs need to be accounted for additionally, higher maximum tolerable costs are possible. For households with an EV and a PV system, maximum tolerable costs range from 211€ (low price scenario) to 413€ (high price scenario). For households with only an EV, maximum tolerable costs range from 50€ (low price scenario) to 316€ (high price scenario). For households with a HP, lower cost savings can be achieved in the DT-flex case, therefore the maximum tolerable costs are also lower, ranging from 84€ (low price scenario) to 111€ (high price scenario). Adding a PV system increases the maximum tolerable costs for HEMS and metering point operation that would be compensated through smart operation to 186€ (low price scenario) and 237€ (high price scenario). An additional BSS reduces the maximum tolerable costs in the low and medium price scenarios[12], but in the high price scenario, maximum tolerable costs increase slightly.

Table 3: Maximum annual tolerable costs for the investment costs for HEMS or both, the investment costs for HEMS and metering point operation costs for smart meters for the SC-flex and DT-flex case, and all electricity price scenarios

| Price scenario | SC-flex case | | | DT-flex case | | |
|---|---|---|---|---|---|---|
| | Low | Medium | High | Low | Medium | High |
| EV | - | - | - | 50 € | 188 € | 316 € |
| EV + PV | 171 € | 183 € | 196 € | 211 € | 310 € | 413 € |
| EV + PV-BSS | 126 € | 136 € | 145 € | 171 € | 291 € | 420 € |
| HP | - | - | - | 84 € | 97 € | 111 € |
| HP + PV | 200 € | 241 € | 282 € | 186 € | 209 € | 237 € |
| HP + PV-BSS | 146 € | 183 € | 220 € | 144 € | 197 € | 255 € |
| EV + HP | - | - | - | 153 € | 330 € | 501 € |
| EV + HP + PV | 354 € | 410 € | 466 € | 390 € | 545 € | 695 € |
| EV + HP + PV-BSS | 281 € | 331 € | 383 € | 324 € | 500 € | 694 € |

Overall, in the DT-flex case, households with only an EV determine the lowest maximum tolerable costs (50€) in the low price scenario. For the medium price scenario, the lowest maximum tolerable costs are defined by households with only a HP (97€). This is also the case for the high price scenario, with the lowest maximum tolerable costs being around 111€ for households with a HP.

---

[12] The reason for this is that households with a BSS already experience lower overall electricity costs even without making use of flexibility. Hence, the additional cost savings gained through flexibility utilization in such households are relatively modest.

In the SC-flex case, households with an EV and a PV-BSS determine the lowest maximum tolerable costs in all price scenarios ranging from 126€ in the low price scenario to 145€ in the high price scenario.

## 4  Discussion

Our analysis shows that rising electricity prices and higher price spreads are increasing the financial incentives for households to utilize the flexibility of their HPs and EVs, both independently and in combination with PV systems and PV-BSSs. With the appropriate enabling technologies (HEMS, smart meter), cost savings, in terms of the annual unit rate costs, can be seen for both the DT-flex and SC-flex cases for all technology combinations. The benefits are highest for households with both an EV and a HP, highlighting the importance of accounting for the interaction of both technologies in one household. In case a dynamic tariff is used (DT-flex case), the comparably largest benefits can be seen for households with an EV if a dynamic tariff is applied and available flexibility is used. Our findings are consistent with those reported in Refs. [10,11]. However, Ref. [12] suggests that households with both an EV and a PV system could achieve higher cost savings by optimizing self-sufficiency than by using a DA-RTP tariff, which contradicts our findings, most likely due to the higher price spreads assumed in our calculations. We find that households with a HP can achieve cost savings when using a dynamic tariff, which is consistent with the results found in Refs. [14–16].

Nevertheless, our results show that using a HEMS to increase self-consumption with a static tariff (SC-flex case) is the best choice for households with a HP. This finding shows the importance of comparing the use of dynamic tariffs not only with the case of no flexibility use at all but also with the case of smart operation of flexible technologies for self-consumption. Furthermore, results show an increase in self-consumption rates for households with an EV or a HP in combination with a PV-system or PV-BSS for both, the DT-flex and the SC-flex case. This could lead to an increase in the specific grid charges for households, as grid costs are allocated to the end-users based on the amount of energy passed through a grid.

Considering the associated costs of HEMS and smart meters, we analyzed the highest allowable costs for both technologies that would still make the utilization of flexibility financially appealing for a minimum of 75% of the households considered. In this context, households equipped with both an EV and a HP can tolerate the highest costs while benefiting the most from their flexibility utilization.

Focusing on the specific technology combinations that dictate this price ceiling, our results indicate that for the SC-flex case, households possessing an EV along with a PV-BSS establish the maximum acceptable costs for the HEMS across all electricity price scenarios. Moreover, these households find dynamic tariffs more financially advantageous. Consequently, the ceiling cost for HEMS in the SC-flex case could be set according to what is tolerable for households with both a HP and a PV system.

For the DT-flex case, households with only an EV present the lowest ceiling costs in the low price scenario. However, in the medium and high price scenarios, this shifts towards households equipped with a HP. Given that those households do not gain additional financial benefits from adopting a dynamic tariff over increasing their self-consumption, the actual expenses for HEMS and metering point operation could be tailored to accommodate households with an EV.

Looking at the effects on household load curves, in the DT-flex case, we observe a shift in households' load to the early morning hours and an increase in peak load when electricity prices are the lowest for households with an EV or a HP. Our analysis revealed that for air/water HPs to respond effectively to dynamic tariffs, there needs to be a larger spread between high and low prices. Otherwise, we observe a load shift predominantly towards midday. This midday load shift is a result of the interaction between the ambient temperature-dependent COP and dynamic prices.

While our study is based on households and electricity prices in Germany, featuring three different electricity price scenarios, the findings can offer broader insights applicable to countries with similar spot market pricing structures and end-user costs. We used the price development from 2021 and 2022 to project higher price levels and larger price spreads, which are also anticipated with the growing share of renewable energies. It is important to note that in reality, electricity prices are influenced by a multitude of factors, such as the increase in renewable energies, the replacement and phase-out of fossil generators, and the electricity market design. Thus, while our analysis can indicate the financial attractiveness of adopting a HEMS or combining a HEMS with dynamic electricity tariffs for households, it does not have the scope to forecast future electricity prices.

Additionally, our study did not account for the investments associated with electric vehicles, heat pumps, or battery storage systems. These technologies are generally acquired by households with the expectation of static tariffs, and potential cost savings through dynamic tariffs are considered post-purchase. Including the investments would broaden the results but is outside the scope of this analysis.

The concept of perfect foresight neglects forecasting errors that would occur in reality and would reduce potential benefits. However, in combination with a rolling horizon within the model EVaTar-building forecasting errors are considered to some extent. Therefore, the analysis shows a more realistic picture than the maximum benefit and provides indications for potential system benefits, if available flexibility is used by the different stakeholders.

Concerning heat pumps, our case study only included air/water heat pumps. Brine/water heat pumps, which exhibit a stable COP not dependent on ambient temperature, may offer more flexibility in response to electricity prices and thus are easier to 'control' via pricing incentives.

We also assumed a constant indoor temperature within the homes. In actual scenarios, households may opt for lower temperatures during periods of elevated energy costs. This aspect deserves further research attention. Lastly, electric vehicles in our model are depicted as having a season-independent energy consumption. In reality, consumption is expected to vary with the seasons, particularly due to heating and cooling needs. Despite this, similar flexible behavior can be expected when exposed to dynamic tariffs.

## 5  Summary and Conclusion

In this study, we analyze the effects of higher electricity prices and larger price spreads based on the development of the day-ahead spot market in Germany from 2019 to 2022 on the financial attractiveness of smart operation of electric vehicles and heat pumps. We address several research gaps by considering the costs for enabling technologies such as home energy management systems and metering point operation, including dynamic tariffs as well as self-consumption and the heterogeneity of households and their flexible assets.

We combine 316 measured household load profiles with different technologies such as electric vehicles, heat pumps, PV rooftop systems, and battery storage systems. The flexibility use is depicted by using a MILP model that minimizes the households' electricity bill and therefore maximizes the households' economic benefits. We analyze flexibility utilization using a home energy management system to increase the self-consumption of PV systems (SC-flex case) and to take advantage of dynamic electricity prices (DT-flex case). We also define three price scenarios by manipulating the day-ahead spot market price time series of 2019 to fit the given average, standard deviation, and min/max values.

Our findings show that dynamic electricity prices based on the day ahead spot market in Germany can offer economic benefits to households equipped with different flexible technologies and therefore incentivize using the flexibility of electric vehicles and heat pumps. To realize these benefits, a certain price level and price spreads are essential to offset the additional costs for HEMS and metering point operation of smart meters.

In our scenarios, an increase in the average electricity price of 15.2 €ct/kWh (+67%) and an increase in the average price spread of 8.9 €ct/kWh (+494%), significantly boosts the share of households opting for the dynamic tariff – from an initial 3.9% to 62.5%. For households with a heat pump, focusing on optimizing self-consumption can yield greater savings than with dynamic tariffs, especially if a PV system or a PV battery storage system is present. Conversely, for households with electric vehicles, dynamic tariffs prove to be more financially attractive than concentrating solely on self-consumption. For households equipped with both an electric vehicle and a heat pump, even higher cost savings are possible when using a dynamic tariff, making flexibility utilization more profitable and compensating for the effort. Our findings suggest that dynamic tariffs can effectively enhance flexibility utilization, but, depending on the available technologies, are not necessarily the best choice financially.

From a market perspective, it can be advantageous for providers of home energy management systems to offer their systems combined for households with both electric vehicles and heat pumps, as this is where the most significant savings potential lies.

It can be assumed that EVs shift their load to the lowest prices of the day (early morning hours) provided the vehicle is available at home during that time. For households equipped with a heat pump, biggest benefits can be created in winter. Here, the co-dependency between prices and the temperature-dependent COP needs to be considered, leading the heat pump to also shift load to midday. Digitization, embodied in HEMS and smart meters, plays a crucial role in harnessing this flexibility, but the costs involved must remain within a certain threshold to maintain financial viability for households. Financial gains from dynamic tariffs are less attractive for these households once the costs of HEMS and smart metering are included, yet their flexibility could be critical in winter due to anticipated simultaneous heating demands.

The cost savings associated with the usage of a home energy management system and/or dynamic tariffs can make technologies such as heat pumps and electric vehicles financially more attractive and help promote a faster transformation[13]. As potential cost savings increase for households with electric vehicles and heat pumps when a PV rooftop system is added, the maximum allowable costs for a home energy management system and metering point operation increase, making the use of flexibility and dynamic tariffs more attractive.

To promote greater flexibility in residential electricity demand, efforts should be focused on expanding rooftop PV systems and accelerating the rollout of smart meters. Reducing the costs associated with smart meter operation could further incentivize dynamic tariff adoption.

With regard to the widely discussed possibility of alleviating grid bottlenecks through dynamic pricing, the results imply that financial incentives need to be set at a sufficiently high level. This is particularly crucial for households with heat pumps, to ensure they respond to dynamic pricing signals rather than solely using their flexibility to enhance self-consumption.

Further research should explore the impact of forecasting errors and the quality of forecasts in home energy management systems. One interesting point would be to compare the quality of forecasts of home energy management systems with forecasts of service providers and the effects and potential cost savings for the customer. Another interesting aspect is the multipurpose use of flexibility as flexibility of households can for example also be utilized for grid stabilization through dynamic grid charges or other control signals from the distribution system operators.

---

[13] For electric vehicles: if we assume a invest of 25.000€ with an interest rate of 2% and a lifetime of 15 years we get an annual annuity of 1.375€. In comparison to that, with an PV-BSS system, yearly savings of 420€ (so roughly 30% of the annual annuity) were shown for the high price scenario; For heat pumps: if we make the same assumptions for the annual annuity of heat pumps, we can compare it to yearly savings of 255€ (around 18% of the annual annuity) for households with an additional PV-BSS system.

# Acknowledgements

Funding: This work was supported by the German Federal Ministry for Economic Affairs and Energy in the context of the research project 'DiMA-Grids' [grant number 03EI6038A].

# Annex

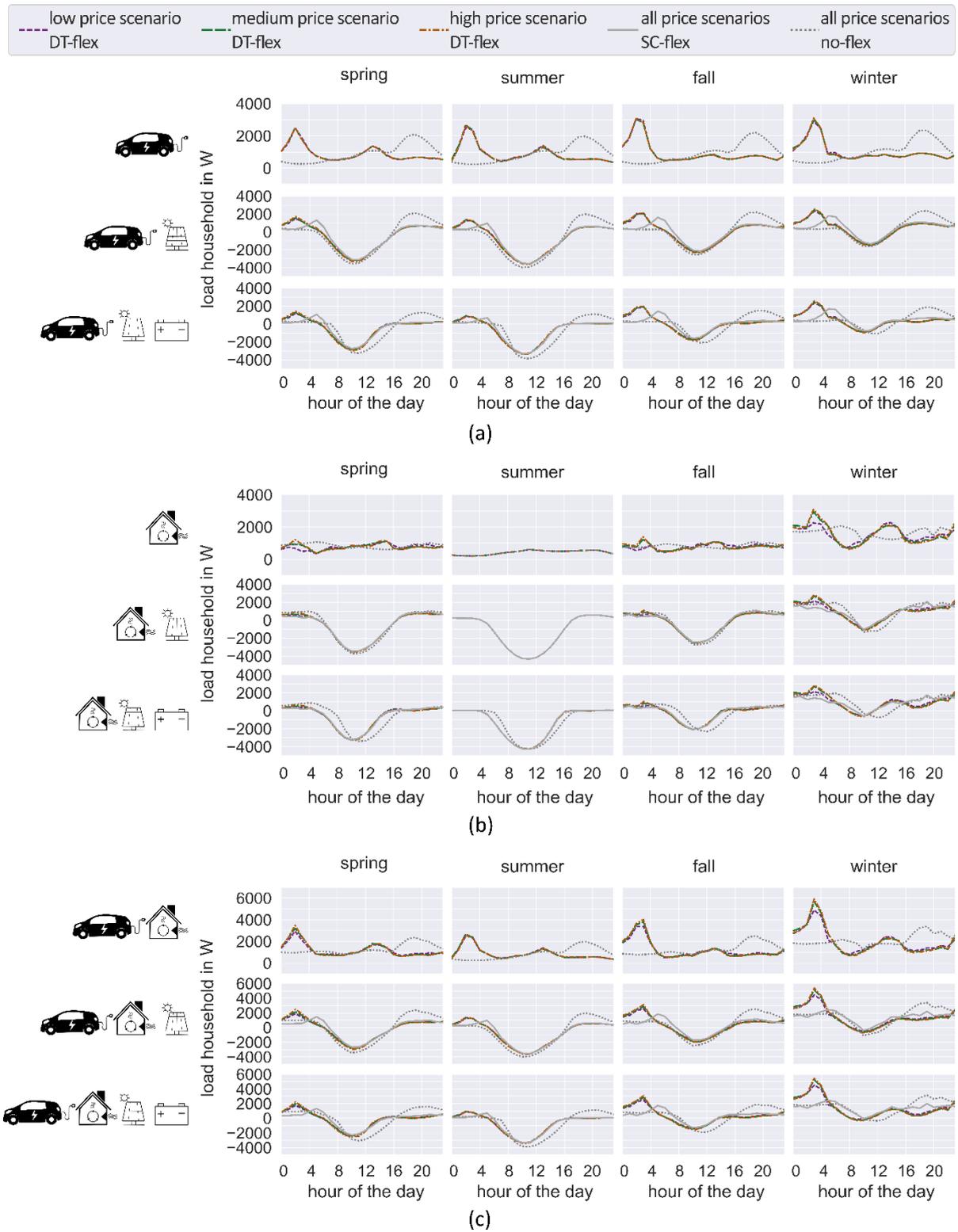

*Figure Annex 1: Average household load per season for each hour of the day including all households considered for the defined cases and scenarios.*

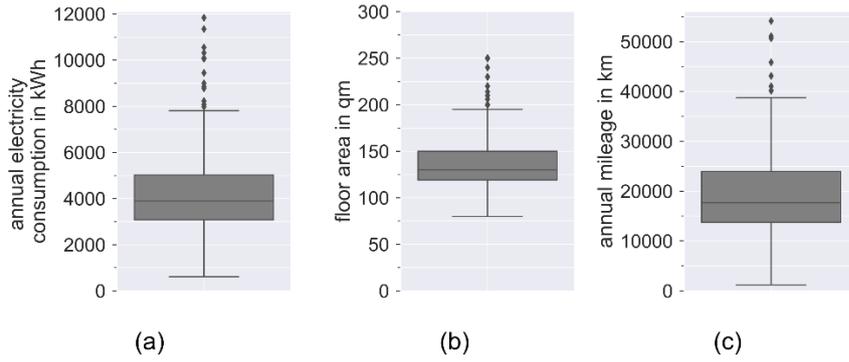

*Figure Annex 2: (a) boxplot of annual electricity consumption for all households considered, (b) boxplot of the heated floor area of all households considered, (c) boxplot of the annual mileage of all EVs considered*

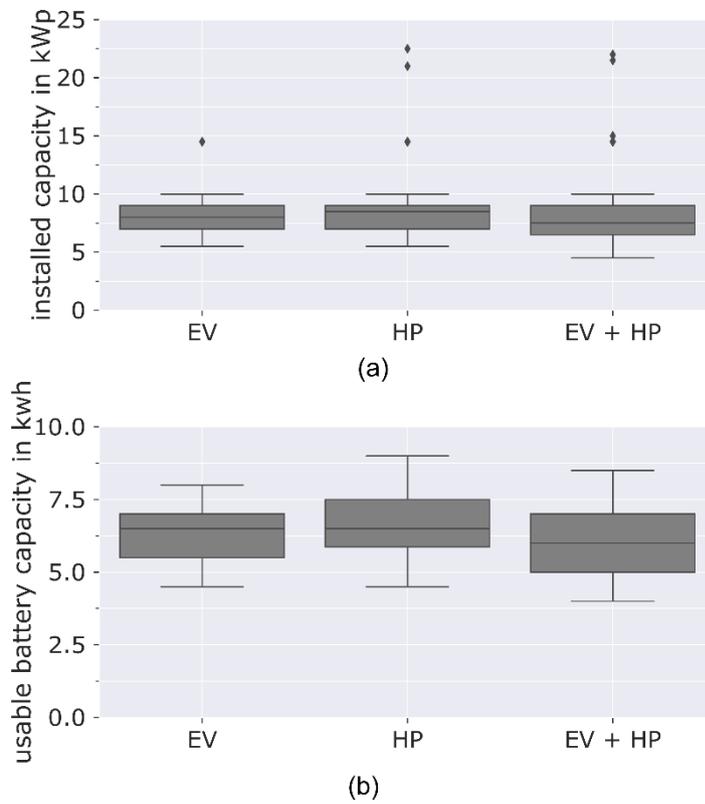

*Figure Annex 3: (a) boxplot of the installed PV capacity for all households with an EV, a HP, or an EV and a HP, (b) boxplot of the usable battery capacity for all households with an EV, a HP, or an EV and a HP*